\documentclass[12pt]{article}
\usepackage{amsmath}
\usepackage{amssymb}

\DeclareMathOperator{\sech}{sech}

\title{
Revisiting an isospectral extension of the Morse potential}
\author{C. Quesne\thanks{e-mail: cquesne@ulb.ac.be}\\
{\small\sl Physique Nucl\'eaire Th\'eorique et Physique Math\'ematique,  Universit\'e Libre de Bruxelles,} \\ 
{\small\sl Campus de la Plaine CP229, Boulevard~du Triomphe, B-1050 Brussels, Belgium}}
\date{ }
\begin{document}
\baselineskip=22pt plus 1pt minus 1pt
\maketitle

\begin{abstract} 
A very simple method is devised to derive a (strictly) isospectral extension of the Morse potential. Furthermore, point canonical transformations are used to transform the latter into quasi-exactly solvable extensions of the radial oscillator and the Coulomb potentials.
\end{abstract}

\noindent
Keywords: quantum mechanics; supersymmetry; quasi-exact solvability

\noindent
PACS Nos.: 03.65.Fd, 03.65.Ge
%
%
\newpage

\section{Introduction}

Extensions of the Morse potential obtained by adding some nonsingular function in such a way that the resulting potential remains exactly solvable (ES) have been carried out in many different ways.\par
%
%
In the context of unbroken supersymmetric quantum mechanics (SUSYQM) \cite{cooper} for shape invariant (SI) potentials \cite{genden} (or in the equivalent framework of backward Darboux transformations \cite{darboux}), one may quote the use of the confluent hypergeometric function and its derivative as building blocks \cite{junker}, the resort to algebraic deformations \cite{gomez04}, or the application of nonuniqueness of factorization \cite{berger, dutta}.\par
%
%
After the introduction of exceptional orthogonal polynomials in the context of Sturm-Liouville theory \cite{gomez10, gomez09} and the realization of their usefulness in constructing new SI extensions of ES potentials in quantum mechanics \cite{cq08a, cq09}, the Morse potential was reexamined by making use of the Darboux-B\"acklund transformation \cite{grandati11} or the prepotential method \cite{ho}. Later on, another family of Morse potential extensions exhibiting an ``enlarged'' SI property was also shown to exist \cite{cq12, grandati12}.\par
%
%
On using still another approach of SI potentials based on expansions in powers of $\hbar$ \cite{bougie10, bougie12}, Bougie, Gangopadhyaya, Mallow, and Rasinariu constructed an isospectral extension of the Morse potential \cite{bougie15}, which was clearly different from the already known ones. Very recently, by performing changes of variable and of parameters, they proved that such a ``new'' extension was actually equivalent to the Scarf II potential, which provided them with a path for going from SI potentials connected with the confluent hypergeometric equation to those related to the hypergeometric one \cite{mallow}.\par
%
%
The purpose of the present work is twofold. First, we will show that instead of the rather complicated use of an expansion in powers of $\hbar$, there exists an alternative and very simple method to derive the isospectrally extended Morse potential or the equivalent Scarf II potential of Refs.~\cite{bougie15, mallow}. Second, we will take advantage of the known point canonical transformations (PCT) connecting the Morse potential to the radial oscillator and the Coulomb potentials \cite{cooper, haymaker, de, cq08b} to map the same extended Morse potential onto quasi-exactly solvable \cite{turbiner, shifman, ushver} radial oscillator and Coulomb extensions, respectively.\par
%
%
\section{Isospectral extension of the Morse potential}

In (first-order) SUSYQM, a general Hamiltonian is written in terms of linear operators
\begin{equation}
  A = \frac{d}{dx} + W(x), \qquad A^{\dagger} = - \frac{d}{dx} + W(x),
\end{equation}
where $W(x)$ is a function of $x$ and some parameters, called the superpotential. In units wherein $2m=\hbar=1$, the starting Hamiltonian is given by
\begin{equation}
  H^{(+)} = A^{\dagger} A = - \frac{d^2}{dx^2} + V^{(+)}(x) - \epsilon,
\end{equation}
where
\begin{equation}
  V^{(+)}(x) = W^2(x) - W'(x) + \epsilon,
\end{equation}
$\epsilon$ is the factorization energy, assumed smaller than or equal to the ground-state energy of $V^{(+)}(x)$, and a prime denotes a derivative with respect to $x$. To $\epsilon$, one can associate a factorization function $\varphi(x)$, which is a nodeless solution of the Schr\"odinger equation
\begin{equation}
  \left(- \frac{d^2}{dx^2} + V^{(+)}(x)\right) \varphi(x) = \epsilon \varphi(x),
\end{equation}
and in terms of which the superpotential can be expressed as $W(x) = - (\log \varphi(x))'$.\par
%
%
To $H^{(+)}$ corresponds a partner Hamiltonian
\begin{equation}
  H^{(-)} = A A^{\dagger} = - \frac{d^2}{dx^2} + V^{(-)}(x) - \epsilon,
\end{equation}
with
\begin{equation}
  V^{(-)}(x) = W^2(x) + W'(x) + \epsilon.
\end{equation}
The pairs of Hamiltonians intertwine with $A$ and $A^{\dagger}$ as $A H^{(+)} = H^{(-)} A$ and $A^{\dagger} H^{(-)} = H^{(+)} A^{\dagger}$.\par
%
%
Here, we restrict ourselves to the case where $\epsilon = E^{(+)}_0$ and $\varphi(x) = \psi^{(+)}_0(x)$ are the ground-state energy and wavefunction of $V^{(+)}(x)$, so that $W(x) = - \left(\log \psi^{(+)}_0(x)\right)'$. The spectra of the two potentials are then related through the relation $E^{(+)}_{n+1} = E^{(-)}_n$, $n=0$, 1, 2, \ldots.\par
%
%
{}Furthermore, we assume that the starting potential is SI, which means that the two partner potentials $V^{(+)}$ and $V^{(-)}$ are similar in shape and differ only in the parameters $a$ that appear in them, i.e.,
\begin{equation}
  V^{(-)}(x,a_0) + g(a_0) = V^{(+)}(x,a_1) + g(a_1),
\end{equation}
or
\begin{equation}
  W^2(x,a_0) + W'(x,a_0) + g(a_0) = W^2(x,a_1) - W'(x,a_1) + g(a_1),  \label{eq:SI}
\end{equation}
where $a_1$ is some function of $a_0$ and $g(a_0)$, $g(a_1)$ do not depend on $x$. In particular, we consider the case of translational (or additive) shape invariance, wherein $a_1$ and $a_0$ only differ by some additive constant. Considering then a set of parameters $a_i$, $i=0$, 1, 2,~\ldots, one can write the eigenvalues of $V^{(+)}(x,a_0)$ as $E^{(+)}_n(a_0) = g(a_n)$, $n=0$, 1, 2,~\ldots.\par
%
%
{}For the Morse potential, in particular,
\begin{align}
  V^{(+)}(x,A) &= V_{A,B}(x)  = B^2 e^{-2x} - B(2A+1) e^{-x}, \\
  W(x,A) &= A - B e^{-x},  \label{eq:W-Morse}
\end{align}
for which $-\infty < x < \infty$ and $A$, $B>0$, one has $a_0 = A$, $a_1 = A-1$, and $g(a_0) = g(A) = -A^2$. This means that Eq.~(\ref{eq:SI}) can be written as
\begin{equation}
  W^2(x,A) + W'(x,A) - A^2 = W^2(x,A-1) - W'(x,A-1) - (A-1)^2 \label{eq:SI-Morse}
\end{equation}
and that the bound-state spectrum of $V^{(+)}$ is given by $E^{(+)}_n = g(A-n) = - (A-n)^2$, $n=0$, 1,~\ldots, $n_{\rm max}<A$.\par
%
%
The purpose of the present section is to construct a general isospectral extension of the Morse potential, i.e., a more general solution of Eq.~(\ref{eq:SI-Morse}) than that given in (\ref{eq:W-Morse}), which, for some limiting values of its parameters, will go back to the latter.\par
%
%
Let us assume a superpotential of the form
\begin{equation}
  W(x,A) = AX_1(x) + X_2(x),  \label{eq:W-Morse-gen}
\end{equation}
where $X_1(x)$ and $X_2(x)$ are two yet undetermined functions of $x$. On inserting this equation in Eq.~(\ref{eq:SI-Morse}), we obtain that $X_1(x)$ and $X_2(x)$ must satisfy the system of differential equations
\begin{align}
  X'_1 + X_1^2 &= 1,  \label{eq:eq1}\\
  X'_2 + X_1 X_2 &= 0.  \label{eq:eq2}
\end{align}
\par
%
%
Equation (\ref{eq:eq1}) is a Riccati equation with a known particular solution $X_1=1$ (corresponding to the starting Morse potential). It is therefore possible to find its general solution by setting \cite{ince}
\begin{equation}
  y = \frac{1}{X_1-1} \qquad \text{or} \qquad X_1 = 1 - \frac{1}{y}.  \label{eq:y}
\end{equation}
The new function $y$ satisfies the first-order differential equation
\begin{equation}
  y' = 2y+1,
\end{equation}
whose general solution can be written as
\begin{equation}
  y = - \frac{1}{2} \left(1+\frac{1}{Q}e^{2x}\right),
\end{equation}
in terms of some constant $Q$. From (\ref{eq:y}), we therefore find that the general solution of the Riccati equation (\ref{eq:eq1}) is
\begin{equation}
  X_1(x) = 1 - \frac{2Q}{e^{2x}+Q}.  \label{eq:X_1}
\end{equation}
\par
%
%
Equation (\ref{eq:eq2}) can now be easily solved for $X_1(x)$ given in (\ref{eq:X_1}). Its general solution depends on another constant, which we set equal to $2P-B$, and is given by
\begin{equation}
  X_2(x) = (2P-B) \frac{e^x}{e^{2x}+Q}.  \label{eq:X_2}
\end{equation}
\par
%
%
With $X_1$ and $X_2$ obtained in (\ref{eq:X_1}) and (\ref{eq:X_2}), respectively, Eq.~(\ref{eq:W-Morse-gen}) becomes
\begin{equation}
  W(x,A) = W_0(x,A) + \phi(x,A). \label{eq:W-Morse-ext}
\end{equation}
where
\begin{equation}
  W_0(x,A) = A - Be^{-x}
\end{equation}
is the starting Morse superpotential (\ref{eq:W-Morse}) and $\phi(x,A)$ is the extension
\begin{equation}
  \phi(x,A) = \frac{2Pe^x - 2AQ + BQe^{-x}}{e^{2x}+Q},
\end{equation}
which goes to zero for vanishing $P$ and $Q$ and coincides with the result obtained in \cite{bougie15} after expanding the superpotential in powers of $\hbar$ and solving the resulting set of partial differential equations.\par
%
%
The extended Morse potential corresponding to (\ref{eq:W-Morse-ext}) is given by
\begin{equation}
  V^{(+)}(x) = W_0^2 + 2W_0\phi + \phi^2 - W_0' - \phi' - A^2
\end{equation}
or
\begin{align}
  V^{(+)}(x) &= V_{A,B,\rm{ext}}(x) \nonumber \\
  &= V_{A,B}(x) + \frac{1}{(1+Q e^{-2x})^2} \{2P(2A+1)e^{-x} \nonumber \\
  &\quad + 4[P(P-B)-A(A+1)Q]e^{-2x} - (2A+1)(2P-3B)Q e^{-3x} \nonumber \\
  &\quad - 2QB^2 e^{-4x} + (2A+1)BQ^2 e^{-5x} - B^2Q^2 e^{-6x}\} \label{eq:V-Morse-ext}
\end{align}
and its partner $V^{(-)}(x)$ is obtained from it by replacing $A$ by $A-1$.\par
%
%
The changes of variable and of parameters
\begin{equation}
  z = x-q, \qquad Q = e^{2q}, \qquad B' = \frac{1}{2}(2P-B)e^{-q}
\end{equation}
transform the superpotential (\ref{eq:W-Morse-ext}) and the potential (\ref{eq:V-Morse-ext}) into the Scarf II superpotential and potential
\begin{align}
  W&= A \tanh z + B' \sech z, \\
  V^{(+)} &= [B^{\prime 2} - A(A+1)] \sech^2 z + B'(2A+1) \sech z \tanh z,
\end{align}
respectively. The inverse transformation allows one to obtain the bound-state wavefunctions of the extended Morse potential (\ref{eq:V-Morse-ext}) from the known ones of the Scarf II potential. The latter can be expressed in terms of either Jacobi polynomials with complex variable and parameters \cite{cooper} or Romanovski polynomials with real variable and parameters (see \cite{cq13} and references quoted therein). On using the latter possibility, we get for the bound-state wavefunctions of (\ref{eq:V-Morse-ext}), with energy $E^{(+)}_n = - (A-n)^2$,  
\begin{align}
  \psi^{(+)}_n(x) &\propto [\sinh(x-q)]^A \exp\left\{-\left(P-\frac{B}{2}\right) e^{-q} \arctan[\sinh(x-q)]
       \right\} \nonumber \\
  &\quad \times R_n^{\left(-(2P-B)e^{-q}, -A+\frac{1}{2}\right)}(\sinh(x-q)), \quad n=0, 1, \ldots, n_{\rm max}
       < A.  \label{eq:ext-Morse-wf}
\end{align}
\par
%
%
\section{Quasi-exactly solvable extension of the radial oscillator potential}
  
It is well known \cite{cooper, cq08b} that the Schr\"odinger equation for the Morse potential
\begin{equation}
  \left[- \frac{d^2}{dx^2} + B^2 e^{-2x} - B(2A+1) e^{-x} - E_n\right] \psi_n(x) = 0, \qquad -\infty<x<\infty,
  \label{eq:SE-Morse}
\end{equation}
with
\begin{equation}
\begin{split}
  E_n &= - (A-n)^2, \\
  \psi_n(x) &\propto \left(2Be^{-x}\right)^{A-n} \exp\left(-Be^{-x}\right) L_n^{(2A-2n)}\left(2Be^{-x}\right),
\end{split}
\end{equation}
and $L_n^{(\alpha)}(y)$ a Laguerre polynomial, can be transformed into that for the radial oscillator potential
\begin{equation}
  \left[- \frac{d^2}{dr^2} + \frac{1}{4}\omega^2 r^2 + \frac{l(l+1)}{r^2} - \tilde{E}_n\right] \tilde{\psi}_n(r)
  = 0, \qquad 0<r<\infty,  \label{eq:SE-rad-osc}
\end{equation}
where
\begin{equation}
\begin{split}
  \tilde{E}_n &= \omega \left(2n+l+\frac{3}{2}\right), \\
  \tilde{\psi}_n(r) &\propto r^{l+1} \exp\left(- \frac{1}{4}\omega r^2\right) L_n^{\left(l+\frac{1}{2}\right)}
      \left(\frac{1}{2}\omega r^2\right)
\end{split}
\end{equation}
by performing the changes of variable $r = e^{-x/2}$ and of function $\tilde{\psi}_n(r) = r^{1/2} \psi_n(x(r))$, as well as multiplying the resulting equation by $4/r^2$. The parameters entering the two equations are linked by the relations
\begin{equation}
  \omega = 4B, \qquad l + \frac{1}{2} = 2(A-n),
\end{equation}
which show that in the resulting equation (\ref{eq:SE-rad-osc}), the energy eigenvalue is fixed to the value $\tilde{E}_n = 4B(2A+1)$.\par
%
%
Instead of the Morse potential in Eq.~(\ref{eq:SE-Morse}), let us now consider its extension $V_{A,B,{\rm ext}}(x)$, given in Eq.~(\ref{eq:V-Morse-ext}). By performing the same transformations as before, we obtain that in Eq.~(\ref{eq:SE-rad-osc}), the radial oscillator potential is replaced by
\begin{align}
  V_{\omega, l, \rm{ext}}(r) &= \frac{1}{4}\omega^2 r^2 + \frac{l(l+1)}{r^2} + \frac{4}{(1+Qr^4)^2}
       \biggl\{[P(4P-\omega)+Q]r^2 - \frac{1}{8}Q\omega^2r^6 \nonumber \\
  &\quad - \frac{1}{16} Q^2\omega^2 r^{10} + \biggl(2n+l+\frac{3}{2}\biggr) \biggl[2P - \biggl(2P
       - \frac{3}{4}\omega\biggr) Qr^4 \nonumber \\
  &\quad + \frac{1}{4} Q^2 \omega r^8\biggr] - \biggl(2n+l+\frac{3}{2}\biggr)^2 Qr^2\biggr\}.
\end{align}
Such an extended radial oscillator has a single known eigenvalue $\tilde{E}_n = \omega\left(2n+l+\frac{3}{2}\right)$ with a corresponding eigenfunction obtained from (\ref{eq:ext-Morse-wf}) in the form
\begin{align}
  \tilde{\psi}_n(r) &\propto r^{2n+l+1} \left(e^{-q}+e^q r^4\right)^{-\frac{1}{2}\left(2n+l+\frac{1}{2}\right)}
        \nonumber \\
  &\quad \times \exp \left[- \left(P-\frac{\omega}{8}\right)e^{-q} \arctan\left(\frac{e^{-q}-e^q r^4}{2r^2}
        \right)\right] \nonumber \\
  &\quad \times R_n^{\left(-\left(2P-\frac{\omega}{4}\right)e^{-q}, -n-l+\frac{1}{4}\right)} \left(\frac{e^{-q}
        - e^q r^4}{2r^2}\right).
\end{align}
\par
%
%
\section{Quasi-exactly solvable extension of the Coulomb potential}

As in the previous section, the Schr\"odinger equation (\ref{eq:SE-Morse}) for the Morse potential can be transformed \cite{cooper, cq08b} into that for the Coulomb potential
\begin{equation}
  \left[- \frac{d^2}{dr^2} - \frac{2Z}{r} + \frac{l(l+1)}{r^2} - \tilde{E}_n\right] \tilde{\psi}_n(r) = 0, \qquad
  0<r<\infty,
\end{equation}
where
\begin{equation}
\begin{split}
  \tilde{E}_n &= - \frac{Z^2}{(n+l+1)^2}, \\
  \tilde{\psi}_n(r) &\propto r^{l+1} \exp\left(- \frac{Zr}{n+l+1}\right) L_n^{(2l+1)}\left(\frac{2Zr}
       {n+l+1}\right),
\end{split}
\end{equation}
by performing the changes of variable $r= e^{-x}$ and of function $\tilde{\psi}_n(r) = r^{-1/2} \psi_n(x(r))$, as well as multiplying the resulting equation by $1/r^2$. The parameters are now linked by
\begin{equation}
  2Z = B(2A+1), \qquad l+\frac{1}{2} = A-n,
\end{equation}
so that in the resulting equation, the energy eigenvalue is fixed to the value $\tilde{E}_n = - B^2$.\par
%
%
The same type of transformations performed for the extended Morse potential (\ref{eq:V-Morse-ext}) leads to the extended Coulomb potential
\begin{align}
  V_{Z,l,{\rm ext}}(r) &= - \frac{2Z}{r} + \frac{l(l+1)}{r} + \frac{1}{(1+Qr^2)^2} \biggl\{4P^2 + Q + 6ZQr
        + 2ZQ^2r^3 \nonumber \\
  &\quad + 4P(n+l+1) \biggl(\frac{1}{r}-Qr\biggr) - 4Q(n+l+1)^2 - \frac{4PZ}{n+l+1} \nonumber \\
  &\quad - \frac{QZ^2}{(n+l+1)^2} r^2 (2+Qr^2)\biggr\},
\end{align}
with a single known eigenvalue $\tilde{E}_n = - Z^2/(n+l+1)^2$ and a corresponding wavefunction obtained from (\ref{eq:ext-Morse-wf}) as
\begin{align}
  \tilde{\psi}_n(r) &\propto r^{n+l+1} \left(e^{-q} + e^q r^2\right)^{-n-l-\frac{1}{2}} \nonumber \\
  &\quad \times \exp\left\{- \left(P - \frac{Z}{2(n+l+1)}\right) e^{-q} \arctan\left(\frac{e^{-q}-e^q r^2}{2r}
       \right)\right\} \nonumber \\
  &\quad \times R_n^{\left(-\left(2P-\frac{Z}{n+l+1}\right)e^{-q}, -n-P\right)} \left(\frac{e^{-q}-e^q r^2}{2r}
       \right). 
\end{align}
\par
%
%
\section{Conclusion}

In the present work, we have derived the extended Morse potential of Ref.~\cite{bougie15} in a much simpler way, thereby also providing an easier pathway from the SI potentials related to the confluent hypergeometric equation to those connected with the hypergeometric one.\par
%
%
{}Furthermore, from such an extended Morse potential, we have obtained by PCT both an extended radial oscillator potential and an extended Coulomb one with one known eigenvalue and one known eigenfunction.\par
%
%
\noindent
Note added in proof. After completion of the present work, Prof.\ A.\ Ramos drew the attention of the author to the fact that an approach similar to that of Sec.~2 had already been used by J.F.\ Cari\~ nena, A.\ Ramos, Rev.\ Math.\ Phys.\ {\bf 12}, 1279 (2000).
%
%
\section*{Acknowledgments}

This work was supported by the Fonds de la Recherche Scientifique - FNRS under Grant Number 4.45.10.08.\par
%
%

\end{document}